\documentclass[
notoc,nohyper]{JHEP} 
\usepackage[psamsfonts]{amsfonts}

\newcommand{\tr}{\mathop{\rm tr}\nolimits}
\newcommand{\str}{\mathop{\rm str}\nolimits}

\newcommand{\SU}{\mathop{\rm SU}}
\newcommand{\U}{\mathop{\rm {}U}}
\newcommand{\rmd}{{\rm d}}
\def\endproof{\hskip0.6em plus0.1em minus0.1em
\setbox0=\null\ht0=5.4pt\dp0=1pt\wd0=5.3pt
\vbox{\hrule height0.8pt
\hbox{\vrule width0.8pt\box0\vrule width0.8pt}
\hrule height0.8pt}}
\newtheorem{theorem}{Theorem}[section]
\newtheorem{lemma}{Lemma}[section]
\newtheorem{conjecture}{Conjecture}[section]
\newtheorem{corollary}{Corollary of conjecture}[section]
\newcommand\fverb{\setbox\pippobox=\hbox\bgroup\verb}
\newcommand\fverbdo{\egroup\medskip\noindent%
			\fbox{\unhbox\pippobox}\ }
\newcommand\fverbit{\egroup\item[\fbox{\unhbox\pippobox}]}
\newbox\pippobox
\title{More about the axial anomaly on the lattice}

\author{Hiroshi Igarashi\\
Department of Mathematical Sciences, Ibaraki University, Mito 310-8512,
Japan\\
E-mail: \email{igarashi@serra.sci.ibaraki.ac.jp}}
\author{Kiyoshi Okuyama\\
Department of Mathematical Sciences, Ibaraki University, Mito 310-8512,
Japan\\
E-mail: \email{okuyama@serra.sci.ibaraki.ac.jp}}
\author{Hiroshi Suzuki\\
Department of Mathematical Sciences, Ibaraki University, Mito 310-8512,
Japan\\
E-mail: \email{hsuzuki@mito.ipc.ibaraki.ac.jp}}
\received{\today} 		
\accepted{\today}		

\preprint{IU-MSTP/49\\\heplat{0206003}}	

\abstract{
We study the axial anomaly defined on a finite-size lattice by using a Dirac
operator which obeys the Ginsparg-Wilson relation. When the gauge group
is~$\U(1)$, we show that the basic structure of axial anomaly on the infinite
lattice, which can be deduced by a cohomological analysis, persists even on
(sufficiently large) finite-size lattices. For non-abelian gauge groups, we
propose a conjecture on a possible form of axial anomaly on the infinite
lattice, which holds to all orders in perturbation theory. With this
conjecture, we show that a structure of the axial anomaly on finite-size
lattices is again basically identical to that on the infinite lattice. Our
analysis with the Ginsparg-Wilson Dirac operator indicates that, in appropriate
frameworks, the basic structure of axial anomaly is quite robust and it
persists even in a system with finite ultraviolet and infrared cutoffs.
}


\begin{document}

\maketitle 

\section{Introduction}

In ref.~\cite{Luscher:1999kn}, L\"uscher pointed out that a cohomological
analysis can be used to determine a basic structure of the axial anomaly in
abelian gauge theories with {\it finite\/} lattice spacings. This work paved a
way to study the axial anomaly in a system with a finite ultraviolet cutoff and
then the technique was applied for various cases~\cite{Fujiwara:2000fi,%
Suzuki:2000ii,Kikukawa:2001kd,Luscher:2000zd,Kikukawa:2001mw}. The crucial
properties which make this analysis possible are the locality, the gauge
invariance and a topological property of the axial anomaly. The axial anomaly
defined by the gauge covariant Dirac operator~\cite{Hasenfratz:1998ft,%
Neuberger:1998fp} which satisfies the Ginsparg-Wilson
relation~\cite{Ginsparg:1982bj}, especially the overlap-Dirac
operator~\cite{Neuberger:1998fp}, in fact possesses the required
properties~\cite{Hasenfratz:1998ri,Luscher:1998pq,Niedermayer:1999bi,%
Hernandez:1999et,Neuberger:2000pz}. A further elaborate analysis with this
recognition finally led to a non-perturbative construction of anomaly-free
abelian chiral gauge theories on the lattice~\cite{Luscher:1999du}.

The cohomological analysis, however, is limited to the case of a lattice with
an infinite size. A direct cohomological analysis for finite-size lattices is
not feasible because (i)~The analysis is based on the lattice Poincar\'e
lemma~\cite{Luscher:1999kn}, which is a lattice analogue of the Poincar\'e
lemma being valid for~${\mathbb R}^d$. When the topology of the lattice is
non-trivial (as is the case for the periodic lattice), one expects a
non-trivial $\rmd$-cohomology on the lattice. (ii)~The cohomology relevant to
an analysis of axial anomaly is a {\it local\/} cohomology, for which the
concept of the locality is vital. The meaning of the locality, however, is not
clear on a lattice with a finite size because a Dirac operator which obeys the
Ginsparg-Wilson relation has to have exponentially decaying
tails~\cite{Horvath:1998cm,Bietenholz:1999dg}.

In this paper, we study the axial anomaly defined on a finite-size lattice by
using the Ginsparg-Wilson Dirac operator. This analysis provides an approach to
the axial anomaly in a system with ultraviolet and infrared cutoffs. As already
noted, a direct generalization of the technique of~ref.~\cite{Luscher:1999kn}
is not feasible. Instead, we point out that it is possible to determine the
structure of axial anomaly using an argument similar to that
of~ref.~\cite{Luscher:1999du} at least in abelian gauge theories. For
non-abelian theories, we propose a conjecture on a possible form of axial
anomaly on the infinite lattice, which is correct within perturbation theory.
Under this conjecture, a similar argument can be applied to non-abelian cases
too. These results indicate that the structure of axial anomaly is quite robust
even with ultraviolet and infrared cutoffs in appropriate formulations (in the
present case, a formulation based on the Ginsparg-Wilson relation). We consider
an even-dimensional lattice~${\mit\Gamma}$ whose size is~$L$,
${\mit\Gamma}=\{\,x\in{\mathbb Z}^d\mid0\leq x_\mu<L\,\}$, and the gauge field
$U(x,\mu)\in G$ ($G$ is the gauge group) is assumed to be periodic
on~${\mit\Gamma}$, $U(x+L\hat\nu,\mu)=U(x,\mu)$.\footnote{$\hat\mu$ denotes the
unit vector in direction~$\mu$.} The lattice spacing~$a$ is set to be unity,
except when the classical continuum limit is considered.

\section{Preliminaries}

The axial anomaly for the Ginsparg-Wilson Dirac operator is defined by (see
for example refs.~\cite{Luscher:1998pq,Niedermayer:1999bi} for the background)
\begin{equation}
   {\cal A}(x)=\tr\gamma_{d+1}\biggl[1-{1\over2}D(x,x)\biggr].
\label{twoxone}
\end{equation}
The kernel of the Dirac operator~$D(x,y)$ satisfies the Ginsparg-Wilson
relation
\begin{equation}
   \gamma_{d+1}D(x,y)+D(x,y)\gamma_{d+1}=D\gamma_{d+1}D(x,y).
\label{twoxtwo}
\end{equation}
The salient feature of~${\cal A}(x)$ is a lattice analogue of the analytic
index theorem~\cite{Hasenfratz:1998ri}
\begin{equation}
   \sum_{x\in{\mit\Gamma}}{\cal A}(x)=n_+-n_-,
\label{twoxthree}
\end{equation}
which follows from the algebraic relation~(\ref{twoxtwo}) alone; here $n_+$
($n_-$) is the number of zero-modes of $\gamma_{d+1}D$ with the positive
(negative) chirality. The index theorem~(\ref{twoxthree}) implies that the
Dirac operator cannot be a smooth function of the gauge field in general,
because the configuration space of lattice gauge field is arcwise connected
and, barring a possibility that $n_+-n_-$ is constant for all configurations,
the integer~$n_+-n_-$ jumps at certain points in the configuration space. A
sufficient condition for the smoothness of the overlap-Dirac
operator~\cite{Neuberger:1998fp} is the
admissibility~\cite{Hernandez:1999et,Neuberger:2000pz}
\begin{equation}
   \|1-U(x,\mu,\nu)\|<\epsilon,\qquad\hbox{for all $x$, $\mu$, $\nu$,}
\label{twoxfour}
\end{equation}
where $U(x,\mu,\nu)$ is the plaquette variable and $\epsilon$ is a constant
smaller than~$(2-\sqrt{2})/d(d-1)$~\cite{Neuberger:2000pz}.\footnote{When the
mass parameter~$m$ in the overlap-Dirac operator is unity, $|m|=1$.} After
imposing this admissibility, the space of allowed gauge field configurations
may have non-trivial topology. This condition also guarantees the locality of
the operator~\cite{Hernandez:1999et,Neuberger:2000pz}
\begin{equation}
   \|D(x,y)\|\leq C(1+\|x-y\|^p)e^{-\|x-y\|/\varrho},
\label{twoxfive}
\end{equation}
where $C$ and $p$ are constants and $\varrho$ is a localization range of the
Dirac operator. In addition to the gauge covariance and the locality of the
Dirac operator, we assume that it has the same transformation law as the
standard Wilson-Dirac operator under discrete symmetries of the lattice
(rotations, reflections, etc.). In particular, we require the translational
invariance, i.e., $D(x,y)$ is identical to $D(x+z,y+z)$ if the gauge field is
shifted at the same time $U(x,\mu)\to U(x+z,\mu)$.

Suppose that we have constructed a Dirac operator on a lattice with the
size~$L$. When $L\to\infty$, $D(x,y)$ is promoted to a Dirac operator on the
infinite lattice $D(x,y)\to D^\infty(x,y)$. This operator also obeys the
Ginsparg-Wilson relation
\begin{equation}
   \gamma_{d+1}D^\infty(x,y)+D^\infty(x,y)\gamma_{d+1}
   =D^\infty\gamma_{d+1}D^\infty(x,y).
\label{twoxsix}
\end{equation}
In what follows, when we compare objects on the finite lattice~${\mit\Gamma}$
and on the infinite lattice, we always take repeated copies of a configuration
of the gauge field on~${\mit\Gamma}$ as the gauge-field configuration on the
infinite lattice. $D(x,x)$ on~${\mit\Gamma}$ and $D^\infty(x,x)$ with the
argument~$x$ restricted to~${\mit\Gamma}$ may somewhat differ because they
have exponentially decaying tails. However we assume that this ``finite size
correction'' is exponentially small,
\begin{equation}
   \|D(x,x)-D^\infty(x,x)\|\leq\kappa L^\nu e^{-L/\varrho},
   \qquad\hbox{for $x\in{\mit\Gamma}$},
\label{twoxseven}
\end{equation}
where $\kappa$ and~$\nu$ are constants. The overlap-Dirac
operator~\cite{Neuberger:1998fp} possesses all the required properties we
assumed above.\footnote{For the overlap-Dirac operator, whose basic building
block is the Wilson-Dirac operator, one can show the relation
$D(x,y)=\sum_{n\in{\mathbb Z}^d}D^\infty(x,y+Ln)$. We thank Yoshio Kikukawa and
Martin L\"uscher for clarifying this point. From this relation and the
locality~(\ref{twoxfive}), one obtains the bound~(\ref{twoxseven}).}

Now, on the infinite lattice, the axial anomaly is given by
\begin{equation}
   {\cal A}^\infty(x)
   =\tr\gamma_{d+1}\biggl[1-{1\over2}D^\infty(x,x)\biggr].
\label{twoxeight}
\end{equation}
This is a topological field in the sense that
\begin{equation}
   \sum_{x\in{\mathbb Z}^d}\delta{\cal A}^\infty(x)=0,
\label{twoxnine}
\end{equation}
where $\delta$ denotes a local variation of the gauge field. This property can
be shown from the Ginsparg-Wilson relation~(\ref{twoxsix}) (see
ref.~\cite{Fujiwara:2000fi} for example). ${\cal A}^\infty(x)$ is thus a local
topological gauge invariant pseudoscalar field.\footnote{A field~$\phi(x)$ is
termed local, when its dependence on the gauge field at a point~$y$ is
exponentially suppressed as $\|x-y\|\to\infty$. For a more precise definition,
see ref.~\cite{Luscher:1999kn}.} When the gauge group is~$\U(1)$, we can then
apply the cohomological analysis~\cite{Luscher:1999kn,Fujiwara:2000fi} to this
quantity. The result
is\footnote{$\partial_\mu$ and~$\partial_\mu^*$ denote the forward and the
backward difference operators respectively:
\begin{equation}
   \partial_\mu f(x)=f(x+\hat\mu)-f(x),\qquad
   \partial_\mu^*f(x)=f(x)-f(x-\hat\mu).
\label{twoxten}
\end{equation}
}
\begin{equation}
   {\cal A}^\infty(x)=q(x)+\partial_\mu^*k_\mu^\infty(x),
\label{twoxeleven}
\end{equation}
where $k_\mu^\infty(x)$ is a local gauge invariant axial vector current (which
is translational invariant) and the topological density~$q(x)$ is given by
\begin{eqnarray}
   q(x)&=&{{\cal N}i^{d/2}\over(4\pi)^{d/2}(d/2)!}\,
   \epsilon_{\mu_1\nu_1\cdots\mu_{d/2}\nu_{d/2}}
   F_{\mu_1\nu_1}(x)F_{\mu_2\nu_2}(x+\hat\mu_1+\hat\nu_1)\cdots
\nonumber\\
   &&\qquad\qquad\qquad\times F_{\mu_{d/2}\nu_{d/2}}
   (x+\hat\mu_1+\hat\nu_1+\cdots+\hat\mu_{d/2-1}+\hat\nu_{d/2-1}),
\label{twoxtwelve}
\end{eqnarray}
with an integer~${\cal N}$. The abelian field strength is defined
by\footnote{For the cohomological argument to apply, the constant~$\epsilon$ in
eq.~(\ref{twoxfour}) has to be smaller than~$1$
and~$|F_{\mu\nu}(x)/i|<\pi/3$~\cite{Luscher:1999kn}.}
\begin{equation}
   F_{\mu\nu}(x)=\ln U(x,\mu,\nu),\qquad
   -\pi<{1\over i}\,F_{\mu\nu}(x)\leq\pi.
\label{twoxthirteen}
\end{equation}
Strictly speaking, the cohomological analysis alone admits a more general form
of~$q(x)$ than~eq.~(\ref{twoxtwelve}); for example,
$\beta_{\mu\nu}F_{\mu\nu}(x)$ with anti-symmetric constants~$\beta_{\mu\nu}$
is also possible. However, since ${\cal A}^\infty(x)$ is a pseudoscalar under
lattice rotations and reflections, one infers that it must be proportional to
the Levi-Civita symbol. Also the numerical coefficient
in~eq.~(\ref{twoxtwelve}) is left undermined in the cohomological analysis. We
can however use a matching with the result in the classical continuum limit;
the integer~${\cal N}$ is given by a sum of chiral charges of massless degrees
of freedom~\cite{Kikukawa:1998pd,Fujikawa:1998if,Adams:1998eg,Suzuki:1998yz,%
Chiu:1998qv,Reisz:1999cm,Frewer:2000ee}.

Note that eq.~(\ref{twoxeleven}) is a statement for finite lattice spacings.
Eq.~(\ref{twoxeleven}) states that, even when the lattice spacing is finite,
the main part of the axial anomaly is given by the topological density~$q(x)$
which has a quite analogous form to the continuum counterpart. On the other
hand, the total divergence term~$\partial_\mu^*k_\mu^\infty(x)$ represents
``lattice artifacts'' in the axial anomaly which depend on the details of the
Dirac operator adopted. Our aim in this paper is to show or argue that the
structure represented by eqs.~(\ref{twoxeleven}) and~(\ref{twoxtwelve})
persists even on finite-size lattices and for general gauge groups~$G$.

\section{Abelian gauge theory $G=\U(1)$}

For the axial anomaly defined on a finite lattice~(\ref{twoxone}), a direct
cohomological analysis is not feasible. Nevertheless, we can show the following
\begin{theorem}
When $G=\U(1)$, if the lattice is sufficiently large compared to the
localization range~$\varrho$ of the Dirac operator, say $L/\varrho\geq n$,
\begin{equation}
   {\cal A}(x)=q(x)+\partial_\mu^*k_\mu(x),
\label{threexone}
\end{equation}
where $k_\mu(x)$ is a gauge invariant periodic current on~${\mit\Gamma}$. The
current~$k_\mu(x)$ moreover satisfies the bound
\begin{equation}
   |k_\mu(x)-k_\mu^\infty(x)|\leq\kappa_1L^{\nu_1}e^{-L/\varrho},
\label{threextwo}
\end{equation}
with constants $\kappa_1$ and~$\nu_1$.
\end{theorem}
We emphasize that, for a sufficiently large~$L$, eq.~(\ref{threexone}) is an
exact statement for the axial anomaly~${\cal A}(x)$. Eq.~(\ref{threextwo})
shows that the current~$k_\mu(x)$ differs from the local
current~$k_\mu^\infty(x)$ defined on the infinite lattice only by an
exponentially small amount. Hence, when the lattice size becomes large compared
to~$\varrho$ and thus when the concept of the locality becomes meaningful, the
current~$k_\mu(x)$ can be regarded as a local current. In this way,
eq.~(\ref{threexone}) shows that the structure of axial anomaly on finite-size
lattices is basically identical to that on the infinite
lattice~(\ref{twoxeleven}). The validity of this theorem has been argued
intuitively by Chiu~\cite{Chiu:1998xf}.

\noindent
\textsc{Proof}. The configuration space of the gauge fields allowed by the
admissibility~(\ref{twoxfour}) consists of many components. Each component is
uniquely characterized~\cite{Luscher:1999du} by the magnetic flux
\begin{equation}
   m_{\mu\nu}
   ={1\over2\pi i}\sum_{s,t=0}^{L-1}F_{\mu\nu}(x+s\hat\mu+t\hat\nu),
\label{threexthree}
\end{equation}
which is an integer. For a configuration with the magnetic flux~$m_{\mu\nu}$,
{}from eq.~(\ref{twoxtwelve}), one has~\cite{Fujiwara:2000wn}
\begin{eqnarray}
   \sum_{x\in{\mit\Gamma}}{\cal A}^\infty(x)
   &=&\sum_{x\in{\mit\Gamma}}q(x)
   ={{\cal N}(-1)^{d/2}\over2^{d/2}(d/2)!}\,
   \epsilon_{\mu_1\nu_1\cdots\mu_{d/2}\nu_{d/2}}
   m_{\mu_1\nu_1}m_{\mu_2\nu_2}\cdots m_{\mu_{d/2}\nu_{d/2}}
\nonumber\\
   &=&{\rm an\ integer},
\label{threexfour}
\end{eqnarray}
where the first equality follows from the translational invariance
of~$k_\mu^\infty(x)$ (namely, $k_\mu^\infty(x)$ is a periodic current
on~${\mit\Gamma}$, when the gauge field is periodic).%
\footnote{Eq.~(\ref{twoxthree}) and theorem~(\ref{threexone}) show that the
index is given by the combination~(\ref{threexfour}) in terms of the magnetic
flux. For the overlap-Dirac operator, this relation has been verified
numerically for $d=2$ and~$d=4$~\cite{Narayanan:ss,Fujiwara:2000hx} and proven
analytically for~$d=2$~\cite{Fujiwara:2000hx}.} Combined with the index
theorem~(\ref{twoxthree}), we see that
$\sum_{x\in{\mit\Gamma}}{\cal A}(x)-\sum_{x\in{\mit\Gamma}}{\cal A}^\infty(x)$
is an integer. This integer is however bounded by an exponentially small
quantity: From the assumed property~(\ref{twoxseven}), one infers that
\begin{equation}
   \sum_{x\in{\mit\Gamma}}[{\cal A}(x)-{\cal A}^\infty(x)]
   \leq\kappa_2L^{\nu_2}e^{-L/\varrho}.
\label{threexfive}
\end{equation}
Therefore, when $L$ is greater than some multiple of~$\varrho$, one has
\begin{equation}
   \sum_{x\in{\mit\Gamma}}[{\cal A}(x)-{\cal A}^\infty(x)]=0.
\label{threexsix}
\end{equation}
For this, we can apply the following
\begin{lemma}
For a periodic field~$c(x)$ on~${\mit\Gamma}$ satisfying
\begin{equation}
   \sum_{x\in{\mit\Gamma}}c(x)=0,
\label{threexseven}
\end{equation}
there exists a periodic current $b_\mu(x)$ which is given by a sum of~$c(y)$,
the precise meaning of which is given in eq.~(\ref{threexnine}) below, such
that
\begin{equation}
   \partial_\mu^*b_\mu(x)=c(x),\qquad
   |b_\mu(x)|\leq2L\max_{x\in{\mit\Gamma}}|c(x)|.
\label{threexeight}
\end{equation}
\end{lemma}
Applying this lemma to eq.~(\ref{threexsix}), we see that there exists a
gauge invariant periodic current~$\Delta k_\mu(x)$ such that
${\cal A}(x)-{\cal A}^\infty(x)=\partial_\mu^*\Delta k_\mu(x)$. This field is
exponentially small, $|\Delta k_\mu(x)|\leq\kappa_1L^{\nu_1}e^{-L/\varrho}$,
thus $k_\mu(x)=k_\mu^\infty(x)+\Delta k_\mu(x)$ which proves the theorem.
\endproof

The assertions of the lemma immediately follow from the explicit construction
of~$b_\mu(x)$ (though this is not unique):
\begin{eqnarray}
   b_\mu(x)&=&{1\over L^{d-\mu}}
   \sum_{y_\mu=0}^{x_\mu}\sum_{y_{\mu+1}=0}^{L-1}\cdots
   \sum_{y_d=0}^{L-1}c(x_1,\ldots,x_{\mu-1},y_\mu,\ldots,y_d)
\nonumber\\
   &&\qquad\qquad-{x_\mu+1\over L^{d-\mu+1}}
   \sum_{y_\mu=0}^{L-1}\cdots
   \sum_{y_d=0}^{L-1}c(x_1,\ldots,x_{\mu-1},y_\mu,\ldots,y_d).
\label{threexnine}
\end{eqnarray}
Note that since $b_\mu(x)$ is given by a sum of the field~$c(x)$, $b_\mu(x)$ is
gauge invariant if so is~$c(x)$.

\section{Non-abelian cases}

For general gauge groups~$G$, a cohomological argument in a non-perturbative
level is not known even on the infinite lattice. Thus we propose a conjecture
on a possible form of~${\cal A}^\infty(x)$:
\begin{conjecture}
For general~$G$,
\begin{equation}
   {\cal A}^\infty(x)=q(x)+\partial_\mu^*k_\mu^\infty(x),
\label{fourxone}
\end{equation}
where $k_\mu^\infty(x)$ is a local gauge invariant axial vector current (which
is translational invariant) and the topological density~$q(x)$ is given by
L\"uscher's topological density~\cite{Luscher:1981zq} and its higher
dimensional extensions.
\end{conjecture}
The explicit expression of L\"uscher's topological density is known only for
$d=2$ and for $d=4$: In our context, it is given by ${\cal N}$ times eq.~(32)
of ref.~\cite{Luscher:1981zq}. We simply assume that the construction can be
pursued for higher dimensional cases.\footnote{For $G=\U(1)$, the
construction of ref.~\cite{Luscher:1981zq} can be generalized to arbitrary
dimensions~\cite{Fujiwara:2000wn}. The equivalence of eq.~(\ref{fourxone})
with eq.~(\ref{twoxeleven}) for $G=\U(1)$ has been
shown~\cite{Fujiwara:2000wn}. See also ref.~\cite{Phillips:us}.} The
construction of ref.~\cite{Luscher:1981zq} does not provide a pseudoscalar
$q(x)$. However, we may always enforce this pseudoscalar property by taking
average over lattice symmetries; we assume that this has been done and $q(x)$
is a pseudoscalar. The topological density has the classical continuum limit
\begin{equation}
   \lim_{a\to0}{1\over a^d}q(x)
   ={{\cal N}i^{d/2}\over(4\pi)^{d/2}(d/2)!}\,
   \epsilon_{\mu_1\nu_1\cdots\mu_{d/2}\nu_{d/2}}
   \tr F_{\mu_1\nu_1}F_{\mu_2\nu_2}\cdots F_{\mu_{d/2}\nu_{d/2}}(x).
\label{fourxtwo}
\end{equation}

At the moment, we cannot prove the above conjecture in the non-perturbative
level. However, we see that the conjecture holds to all orders in perturbation
theory; the following theorem guarantees that a gauge invariant topological
field is unique (up to a total divergence) under certain conditions:
\begin{theorem}
Let $p(x)$ be a local gauge invariant pseudoscalar field (which is
translational invariant) on the infinite lattice whose dependences on the
lattice spacing~$a$ arise only though the gauge field.\footnote{Recall that in
the classical continuum limit the gauge potential is introduced as
$U(x,\mu)={\cal P}\exp[a\int_0^1\rmd t\,A_\mu(x+(1-t)a\hat\mu)]$ where $a$~is
the lattice spacing.} If it is topological
\begin{equation}
   \sum_{x\in{\mathbb R}^d}\delta p(x)=0,
\label{fourxthree}
\end{equation}
and the classical continuum limit $\lim_{a\to0}p(x)/a^d$ vanishes, then to all
orders in perturbation theory,
\begin{equation}
   p(x)=\partial_\mu^*\ell_\mu(x),
\label{fourxfour}
\end{equation}
where $\ell_\mu(x)$ is a local gauge invariant axial vector current.
\end{theorem}

\noindent
\textsc{Proof}. Our proof is rather similar to the cohomological argument of
ref.~\cite{Luscher:2000zd}. We expand $p(x)$ with respect to the bare gauge
coupling constant~$g_0$ introduced by $U(x,\mu)=e^{g_0A_\mu(x)}$
\begin{eqnarray}
   &&p(x)=\sum_{k=1}^\infty p^{(k)}(x),
\nonumber\\
   &&p^{(k)}(x)={g_0^k\over k!}\sum_{y_1,\ldots,y_k}
   p^{(k)}(x,y_1,\ldots,y_k)_{\mu_1\cdots\mu_k}^{a_1\cdots a_k}
   A_{\mu_1}^{a_1}(y_1)\cdots A_{\mu_k}^{a_k}(y_k),
\label{fourxfive}
\end{eqnarray}
where $A_\mu(x)=A_\mu^a(x)T^a$.

First consider $p^{(1)}(x)$. Since $p(x)$ is gauge invariant, $p^{(1)}(x)$ is
invariant under the linearized gauge transformation
\begin{equation}
   A_\mu(x)\to A_\mu(x)+\partial_\mu\omega(x),
\label{fourxsix}
\end{equation}
and also under the constant gauge transformation
\begin{equation}
   A_\mu(x)\to A_\mu(x)+[\omega,A_\mu(x)].
\label{fourxseven}
\end{equation}
Moreover, since $p^{(1)}(x)$ is a local topological pseudoscalar field and
eq.~(\ref{fourxsix}) is the gauge transformation in abelian theory, one can
invoke the cohomological analysis in abelian theory. The result is
\begin{equation}
   p^{(1)}(x)=\partial_\mu^*\ell_\mu^{(1)}(x),\qquad
   \ell_\mu^{(1)}(x)=g_0\sum_y\ell_\mu^{(1)}(x,y)_\nu^aA_\nu^a(y).
\label{fourxeight}
\end{equation}
The local axial vector current~$\ell_\mu^{(1)}(x)$ is invariant under
eqs.~(\ref{fourxsix}) and~(\ref{fourxseven}). A key observation is that, from
$\ell_\mu^{(1)}(x)$, one can construct a field~$\widehat\ell_\mu^{(1)}(x)$ such
that it is invariant under the original non-abelian gauge transformation and
its lowest-order $O(g_0)$ term coincides
with~$\ell_\mu^{(1)}(x)$. This can be accomplished by substituting the gauge
potential~$A_\mu^a(y)$ in eq.~(\ref{fourxeight}) by the
expression~\cite{Luscher:2000zd}
\begin{equation}
   \widehat A_\mu^a(x,y)={2\over g_0}
   \tr\Bigl\{T^a\Bigl[1-W(x,y)U(y,\mu)W(x,y+\hat\mu)^{-1}\Bigr]\Bigr\},
\label{fourxnine}
\end{equation}
where $W(x,y)$ is the ordered product of the link variables from~$y$ to~$x$
along the shortest path that goes first in direction~$1$, then direction~$2$,
and so on. Note that $\widehat A_\mu^a(x,y)$ behaves gauge covariantly under
the original non-abelian gauge transformation. Thus the resulting expression,
\begin{equation}
   \widehat\ell_\mu^{(1)}(x)
   =g_0\sum_y\ell_\mu^{(1)}(x,y)_\nu^a\widehat A_\nu^a(x,y),
\label{fourxten}
\end{equation}
is invariant under the non-abelian gauge transformation due to the invariance
of~$\ell_\mu^{(1)}(x)$ under~eq.~(\ref{fourxseven}). Moreover, since
\begin{equation}
   \widehat A_\mu(x,y)=A_\mu(y)+\partial_\mu^y\omega(x,y)+O(g_0),
\label{fourxeleven}
\end{equation}
with $\omega(x,y)$ the oriented line sum of the gauge potential from $y$
to~$x$, the invariance under~eq.~(\ref{fourxsix}) implies that
$\widehat\ell_\mu^{(1)}(x)=\ell_\mu^{(1)}(x)+O(g_0^2)$.\footnote{The
current~$\widehat\ell_\mu^{(1)}(x)$ so constructed is not an axial vector under
the lattice symmetries. However we can always enforce this by taking average
over lattice symmetries.} Using~$\widehat\ell_\mu^{(1)}(x)$, we may define a
local gauge invariant pseudoscalar field
\begin{equation}
   p(x)-\partial_\mu^*\widehat\ell_\mu^{(1)}(x).
\label{fourxtwelve}
\end{equation}
which has identical properties with~$p(x)$ except that it starts with
$O(g_0^2)$~term. Thus we can repeat the above argument from
eq.~(\ref{fourxfive}) for the field~(\ref{fourxtwelve}). This time, however,
the perturbation series analogous to eq.~(\ref{fourxfive}) starts from $k=2$.

In this way, we repeat the steps from eq.~(\ref{fourxfive})
to~eq.~(\ref{fourxtwelve}) by eliminating the lowest-order term of the
topological field until the first order term becomes $O(g_0^{d/2})$; here a new
situation arises. The cohomological analysis (with the fact that it is a
pseudoscalar) tells that
\begin{eqnarray}
   &&c^{a_1\cdots a_{d/2}}\epsilon_{\mu_1\nu_1\cdots\mu_{d/2}\nu_{d/2}}
   F_{\mu_1\nu_1}^{a_1}(x)F_{\mu_2\nu_2}^{a_2}(x+\hat\mu_1+\hat\nu_1)\cdots
\nonumber\\
   &&\qquad\qquad\qquad\qquad\quad
   \times F_{\mu_{d/2}\nu_{d/2}}^{a_{d/2}}
   (x+\hat\mu_1+\hat\nu_1+\cdots+\hat\mu_{d/2-1}+\hat\nu_{d/2-1})
\nonumber\\
   &&+\partial_\mu^*\ell_\mu^{(d/2)},
\label{fourxthirteen}
\end{eqnarray}
[$F_{\mu\nu}^a(x)=\partial_\mu A_\nu^a(x)-\partial_\nu A_\mu^a(x)$ denotes
the linearized field strength] is a possible form of
$p(x)-\sum_{k=1}^{d/2-1}\partial_\mu^*\widehat\ell_\mu^{(k)}(x)$. However,
since the continuum limit of $p(x)$, $\lim_{a\to0}p(x)/a^d$ vanishes, we infer
that the constants $c^{a_1\cdots a_{d/2}}$ vanish, $c^{a_1\cdots a_{d/2}}=0$.
Thus we again have a total divergence. Further repeating the above procedure,
we finally establish
$p(x)=\partial_\mu^*\sum_{k=1}^\infty\widehat\ell_\mu^{(k)}(x)$.
\endproof

Going back to eq.~(\ref{fourxone}), we note that both ${\cal A}^\infty(x)$
and~$q(x)$ are a local gauge invariant topological pseudoscalar field
(for the latter, those properties follow from the construction
of~$q(x)$~\cite{Luscher:1981zq}). Moreover, they have the same classical
continuum limit~(\ref{fourxtwo}). Thus, applying theorem~4.1 to
${\cal A}^\infty(x)-q(x)$, we see that the conjecture holds to all orders in
perturbation theory.

Now, in the proof of theorem~3.1 in abelian theory, every steps are valid even
for non-abelian theories, except for the crucial relation~(\ref{threexfour}),
namely $\sum_{x\in{\mit\Gamma}}{\cal A}^\infty(x)$ is an integer. With our
conjecture~4.1 for non-abelian cases, this last condition is also satisfied;
$\sum_{x\in{\mit\Gamma}}q(x)$ is L\"uscher's topological charge on a periodic
lattice which is an integer. So, repeating the proof for theorem~3.1, we have
\begin{corollary}
For general $G$, if the lattice is sufficiently large compared to the
localization range~$\varrho$ of the Dirac operator, say $L/\varrho\geq n$,
\begin{equation}
   {\cal A}(x)=q(x)+\partial_\mu^*k_\mu(x),
\label{fourxfourteen}
\end{equation}
where $k_\mu(x)$ is a gauge invariant periodic current on~${\mit\Gamma}$. The
current~$k_\mu(x)$ moreover satisfies the bound
\begin{equation}
   |k_\mu(x)-k_\mu^\infty(x)|\leq\kappa_1L^{\nu_1}e^{-L/\varrho},
\label{fourxfifteen}
\end{equation}
with constants $\kappa_1$ and~$\nu_1$. The topological density~$q(x)$ is given
by L\"uscher's topological density~\cite{Luscher:1981zq} and its higher
dimensional extensions.
\end{corollary}
This corollary states that the basic structure of axial anomaly on finite
lattices is identical that on the infinite lattice. Summing
eq.~(\ref{fourxfourteen}) over the lattice~${\mit\Gamma}$, one has an equality
between the index of the Dirac operator~(\ref{twoxthree}) and the
geometrically-defined lattice topological charge~\cite{Luscher:1981zq}. This
equivalence (``lattice index theorem'') has been thought to be true for long
time since the analyses in refs.~\cite{Narayanan:ss,Narayanan:1997sa}. Our
argument provides a further support for this equivalence.

\section{Conclusion}
In this paper, we have studied the axial anomaly defined on a finite-size
lattice by using a Ginsparg-Wilson Dirac operator. For $G=\U(1)$, we show that
the basic structure of axial anomaly on the infinite lattice, which has a quite
analogous form to the continuum counterpart, persists even on a sufficiently
large finite-size lattices. For general~$G$, we conjectured that the axial
anomaly on the infinite lattice is basically given by L\"uscher's topological
density; actually this holds to all orders in perturbation theory. With this
conjecture, we showed that this structure again persists even on finite-size
lattices. Since L\"uscher's topological density is a geometrically natural
definition of the Chern form in lattice gauge theory (note that it is
proportional to $\str T^{a_1}\cdots T^{a_{d/2}}$), our analysis indicates that
the basic structure of axial anomaly in continuum theory is quite robust and it
persists even in a system with finite ultraviolet and infrared cutoffs. Of
course, we indicated this persistency only in a framework with the
Ginsparg-Wilson relation. To understand precise conditions on the formulation
for this persistency to hold is an interesting open question; for example, one
may enlarge the set of formulations by using the generalized Ginsparg-Wilson
relation~\cite{Fujikawa:2000my}.

In the gauge invariant lattice formulation of abelian chiral gauge
theories~\cite{Luscher:1999du}, a knowledge on the structure of $\U(1)$ gauge
anomaly on finite-size lattices was of crucial importance. Recalling this fact,
we believe that our analyses will be useful in extending the construction of
ref.~\cite{Luscher:1999du} to non-abelian gauge theories.

H.S. would like to thank Takanori Fujiwara, Takahiro Fukui, Yoshio Kikukawa and
Martin L\"uscher for valuable discussions. We are grateful to Kazuo Fujikawa
for a careful reading of the manuscript.

\listoftables		
\listoffigures		

\end{document}